\documentclass[12pt,a4paper,aps]{article}
\usepackage{a4}
\usepackage{epsfig}
\usepackage[hmargin=2cm,vmargin=2cm,nohead]{geometry}
\begin{document}
\par
\title{The Future of Nuclear Energy: Facts and Fiction \\ 
Chapter III: \\
How (un)reliable are the Red Book Uranium Resource Data?}

\author{
Michael Dittmar\thanks{e-mail:Michael.Dittmar@cern.ch},\\
Institute of Particle Physics,\\ 
ETH, 8093 Zurich, Switzerland}
\maketitle

\begin{abstract}
For more than 40 years, the Nuclear Energy Agency of the OECD countries and the 
International Atomic Energy Administration of the United Nations have published a biannual document
with the title ``Uranium Resources, Production and Demand". This book, known as the ``Red Book", 
summarizes data about the actual and near future nuclear energy situation and 
presents the accumulated worldwide knowledge about the existing and expected uranium resources.
These data are widely believed to provide an accurate and solid basis for future decisions about nuclear energy. 
Unfortunately, as it is demonstrated in this paper, they do not.

The conventional worldwide uranium resources are estimated by the authors of the Red Book 
as 5.5 million tons. Out of these, 3.3 million tons are assigned to the reasonable assured category and 
2.2 million tons are associated with the not yet discovered but assumed 
to exist inferred resources. Our analysis shows that neither the 3.3 million tons of ``assured" resources 
nor the 2.2 million tons of inferred resources are justified by the Red Book data and that the 
actual known exploitable resources are probably much smaller. 

Despite many shortcomings of the uranium resource data, 
some interesting and valuable information can be extracted from the Red Book.
Perhaps most importantly, the Red Book resource data can be used to test 
the {\bf ``economic-geological hypothesis"}, 
which claims for example that a doubling of uranium price will increase the amount 
exploitable uranium resources by an even larger factor. 
The relations between the uranium resources claimed for the different resource categories and their associated cost estimates are found to be in clear contradiction with this hypothesis.
\end{abstract}


\newpage
\section{Introduction}
Policy makers almost never discuss uranium resources and many other important resource issues 
in public.  
One reason seems to be that most energy resources are still considered to be {\bf ``no-problem"} and thus a taboo 
topic for worldwide policy makers and their economic or academic advisors. 

However the recent price explosions for oil and related media headlines seem to indicate some attitude 
change with respect to the oil resource situation. In fact, more and more people start to pay attention to questions 
of geological and technological limits of oil extraction capacities. This has resulted in the wish 
to obtain accurate oil and gas resource data, especially from the OPEC countries\cite{G8oil}.

In contrast, the uranium resources appear to be accurately documented 
in the ``Red Book: Uranium Recources, Production and Demand". In this book,  
updated every two years,  the IAEA (International Atomic Energy Agency) from the United Nations 
and the NEA (Nuclear Energy Agency) from the OECD countries have presented for more than 40 years \cite{40years}, their 
collective knowledge about uranium resources and its use for civilian nuclear energy.  
The latest update, the  2007 edition, was published in early June 2008\cite{RB07}. 
This book provides more than 400 pages of detailed information about uranium resources in a large number of countries.
A long history of reporting worldwide uranium resource data with a precision 
between 1/1000 and 1/10000 is believed to demonstrate that  
reliable resource data are presented. The findings of the  Red Book 2007 edition were presented for example in the 
NEA press communiqu\'e\cite{NEA2007press} with (quote): \\

\noindent
{\it ``There is enough uranium known to exist to fuel the world's fleet of nuclear reactors at current consumption rates for at least a century, according to the latest edition of the world reference on uranium published today.
{\bf Uranium 2007: Resources, Production and Demand}, also known as the Red Book, estimates the identified amount of conventional uranium resources which can be mined for less than USD 130/kg* to be about 5.5 million tonnes, up from the 4.7 million tonnes reported in 2005. Undiscovered resources, i.e. uranium deposits that can be expected to be found based on the geological characteristics of already discovered resources, have also risen to 10.5 million tonnes. This is an increase of 0.5 million tonnes compared to the previous edition of the report. The increases are due to both new discoveries and re-evaluations of known resources, encouraged by higher prices. \\
(*On 26 May 2008, the spot price for uranium was USD 156/kg.) \\ 
}

After reading such a declaration, most people will obviously assume that the uranium supply 
situation is safe. Why should one even bother to look into the accumulated uranium data 
or doubt these well respected international organizations with their large scientific staff?   
As a consequence of this attitude, individuals and organizations with different philosophical views about nuclear 
energy almost never question the objectivity and precision of these data\cite{RARuse}.

Unfortunately, as will be shown in the following, the Red Book uranium 
resource data do not measure up to the pretended standards of accuracy. 

In this paper, the Chapter III of ``The Future of Nuclear Energy" \cite{chapter12}, we analyze 
the uranium resource data given in the Red Book 2007\cite{RB07}.
First we present and discuss the overall worldwide uranium resource data and their evolution in section 2. 
In order to investigate the basis for these data, 
the uranium resource data for the 10 countries with more 
than 100000 tons of reasonably assured resources (RAR) are analyzed in 
section 3. Combined, these 10 countries represent about 80\% of the world's total RAR and 
95\% of the economically most interesting RAR cost category, $<$ 40 dollars/kg.
As will be demonstrated in detail, the Red Book 2007 uranium resource data often show 
amazing changes with respect to previous Red Book editions,  some of these individual country resource changes 
appear to be totally unbelievable.

In the final section 4, the long term uranium supply situation and its consequences for the 
future of conventional nuclear fission power plants will be summarized.

\section{Worldwide uranium resources and their evolution}

As highlighted already in chapter I and II of this report, the authors of the Red Book 
do not ignore the possibility that ``uranium supply shortfalls could develop if production facilities are not implemented in a timely manner". However, the world media have essentially only transmitted the statement that 
 ``the identified conventional uranium resources have increased 
from 4.7 million tons in the previous report to 5.5 million tons".

\noindent
In the following we will analyze this apparent 20\% increase in conventional uranium resources in detail. 
In order to do this we start with the methodology on how the authors of the Red Book obtain their data and  
present the definitions of the different uranium resources categories. 

\subsection{Red Book methodology, resource categories and extraction costs}

The authors of the Red Book describe the content and the methodology to obtain the relevant data 
in their own words as (Quote)\cite{RBpreface}:  \\

\noindent
{\it ``The Red Book features a comprehensive assessment of current uranium supply and demand and
projections to the year 2030. The basis of this assessment is a comparison of uranium resource
estimates (according to categories of geological certainty and production cost) and mine production
capability with anticipated uranium requirements arising from projections of installed nuclear
capacity. In cases where longer-term projections of installed nuclear capacity were not provided by
national authorities, projected demand figures were developed with input from expert authorities... \\
The Red Book also includes a compilation and evaluation of previously published
data on unconventional uranium resources... \\
This publication has been prepared on the basis of data obtained through questionnaires sent by
the NEA to OECD member countries (19 countries responded) and by the IAEA for those states that
are not OECD member countries (21 countries responded and one country report was prepared by the
IAEA Secretariat). The opinions expressed in Parts I and II do not necessarily reflect the position of
the member countries or international organisations concerned. This report is published on the
responsibility of the OECD Secretary-General."} \\

In Appendix 2 of the Red Book, a list of reporting organizations and contact persons is given for 
a large number of countries\cite{RBappendix2}. This list indicates that 
uranium resource data are a compilation of data from the different government agencies, sometimes supplemented 
by the data from private transnational mining companies.
As large national and private interests are involved, the objectivity and the accuracy of the presented data is 
certainly not assured.
Thus, the resource data do not represent the results from an accurate scientific analysis of geological data.  
Unfortunately, such possible shortcomings of these 
resource estimates and possible large uncertainties are not mentioned in the Red Book. 

However, in absence of better data and in line with the required political consent from many countries, 
it seems that the editors of the Red Book try to encourage the different countries to provide useful and 
comparable resource data. As a result,  using the US dollar as a universal standard,
consistent categories for uranium resources are defined. 

Unfortunately a few comments, presented in the Appendix 4 \cite{RBappendix4} seem to indicate that 
the Red Book resource data are not as accurate as otherwise stated. \\

\noindent
For example it is written that:
\begin{itemize} 
\item {\it ``The categories are defined according to a believed level of confidence". } \\
But associated probabilities for the believed existence of the resources are not quantified.
\item
{\it ``The resource categories are defined in terms of the 
uranium recovery costs at the ore processing plant."} \\ 
But no explanation on how this cost should have been calculated 
for ``non existing ore processing plants" in ``not yet known 
environments" is given. We predict that such estimates will remain a mystery for a long time. 
\item 
{\it ``It is not intended that the cost categories should follow 
(undefined) fluctuations in market conditions" \cite{RBuranmarket}.}
This can only mean that cost estimates have been done independently
from the mining costs. Not everybody will agree that the increased mining costs of the past 
few years, related among other things to the energy costs and in particular to the 
oil price, are just simple ``market fluctuations".     
\end{itemize}

In summary, the used methodology leaves some ``freedom" on how the 
correspondents from the different countries should present their resource data.  
This ``freedom" could explain some large RAR resource changes found for different countries and from 
the subsequent Red Book editions. \\




The uranium resources are separated into {\bf ``conventional" and ``unconventional"  resources}. 
The conventional resources are 
divided into {\bf Reasonably Assured Resources (RAR)} and the believed to exist {\bf Inferred Resources (IR)}. 
The RAR and IR categories are further subdivided according to the assumed exploitation cost in US dollars.  
These cost categories are given as $<$ 40 dollars/kg, 
$<$ 80 dollars/kg and $<$ 130 dollars/kg. 

The unconventional resources are split into {\bf ``Undiscovered Resources (UR)"}, further separated into 
{\bf ``Undiscovered Prognosticated Resources (UPR)"}, with assumed cost ranges of 
$<$ 80 dollars/kg and $<$ 130 dollars/kg,  
and the {\bf ``Undiscovered Speculative Resources (USR)"}. 
The USR numbers are given for an estimated exploitation cost of \\ 
$<$ 130 dollars/kg and also for the category with an unknown cost.

For the purpose of this analysis, the data from the inclusive ``$<$ x dollars/kg" categories are used 
to calculate the sometimes more informative 
exclusive resource data with extraction costs between 40-80 dollars/kg and 80-130 dollars/kg.

A critical reader of the Red Book will express doubts about the data quality, when  
roughly known numbers are given with an unbelievable precision of 0.1\% or better. 
In this respect it seems to be an ironic mistake that the best known numbers in the RAR categories are  
given with an accuracy of 1/1000 but the speculative IR and UR categories are presented with an accuracy 
of 1/10000\footnote{At least some progress has been made since the Red Book edition 2005, 
when the claimed accuracy was presented with an accuracy of 1 part per million.}. 
Despite such pseudo-precise data, names like ``Undiscovered Resources" and 
``Undiscovered Speculative Resources", should not give a high level of confidence 
in the accuracy of the associated amount of uranium.

A more accurate methodology would include effects from changes 
in the uranium mining technology and its related costs, the quality of the mining equipment, the oil price, salaries 
and the often ignored huge costs to repair environmental damages following the past uranium exploration.   
In addition, detailed information should be provided on (1) how variations of the dollar 
exchange rate have modified the resource data, and (2) how past uranium extractions have been
taken into account. 

We leave it to the reader to reflect on the following question: \\
``Can the existing Red Book methodology result in accurate estimates 
for the discovered and undiscovered uranium resources?" 

\subsection{The economic-geological hypothesis about uranium resources}

According to geological estimates one knows that 
uranium is not a particularly rare metal. Expressed in the words of the relevant WNA document 
one reads (quote)\cite{WNAinf75}: \\
 
{\it ``Uranium is a relatively common metal, found in rocks and seawater.  
Economic concentrations of it are not uncommon." }

Table 1 shows uranium or grade concentrations for different minerals 
in the earth crust and in sea water and in parts per million (ppm). 

{
\begin{table}[h]
\begin{tabular}{|c|c|c|}
\hline
uranium content of:            &  concentration [ppm U]    &  uranium / ton  \\
\hline
Very high-grade-ore  20\% U (Canada)          &  200 000 ppm U             & 200 kg/ton  \\
High-grade-ore  2\% U           &  20 000 ppm U             & 20 kg/ton  \\
Low -grade-ore  0.1\% U         &   1 000 ppm U             &  1 kg/ton  \\
Very low -grade-ore$^{*}$ (Namibia) 0.01\% U         &   100 ppm U             &  0.1 kg/ton  \\

\hline
Granite                        &       4-5 ppm U             & 0.004 kg/ton \\ 
Sedimentary Rock               &       2 ppm U             & 0.002 kg/ton \\ 
Earth continental crust (average)&     2.8 ppm U           & 0.003 kg/ton \\
\hline
Sea water                &       0.003 ppm U         & 0.000003 kg/ton \\
\hline
\end{tabular}\vspace{0.3cm} 
\caption{The numbers are taken from the August 2009 version of the WNA information paper 
``Supply of uranium"~\cite{WNAinf75}. 
The $^{*}$  in the WNA document 
is associated with very low grade uranium mining from the Rossing mine in Namibia and the 
(quote): {\it ``If uranium is at low levels in rock or sands (certainly less than 1000 ppm), 
it needs to be in a form which is easily separated for those concentrations to be called ``ore" - that is, implying that the uranium can be recovered economically.  This means that it need to be in a mineral form that can easily be dissolved by sulfuric acid or sodium carbonate leaching."}.  
}
\end{table}
}

It is generally accepted that the product of the uranium concentration and the 
total amount of uranium which exist in this concentration and in the earth's crust will 
increase by a large factor if the concentration decreases. 
K. S. Deffeyes and I. D. MacGregor have estimated \cite{uraniumsum} 
on rather generally accepted geological methods that 
this trend must continue until the average uranium concentration of 2.8 ppm is reached.   
According to Deffeyes and MacGregor one can expect 
approximately that the amount of extractable uranium will increase by a factor of perhaps 
300, if one can exploit each tenfold decrease in ore grade.  They added the usually ignored statement 
that quote (words in brackets are added by us): \\
{\it ``no rigorous statistical basis exists for expecting a log-normal distribution and  
that this is just an approximate argument as the enormously complex range of geochemical behavior of 
uranium and its wide variety of different (chemical?) binds (determine?) the economic deposit."} 

It is thus important to keep in mind that {\bf resource calculations, 
based on the above methods possibly ignore other important factors which limit the amount of eventually extractable uranium}. 
For example one can imagine that hypothetical super high concentration kg rocks 
exist a few hundred meters deep. However, if these rocks are isolated from each other 
and from any other interesting mineral concentration, 
it can be assumed that sizable amounts of these rocks will essentially never be extracted. 
Thus, in addition to the average ore concentration, 
one finds that the uranium amount in this concentration, its chemical composition and 
its surrounding must play an important role for a potential extraction and the corresponding 
energy cost. 

A consequence of this hypothesis is that, no matter what growth scenario is assumed, 
sufficient uranium resources exist {\bf in theory and if} the extraction cost are allowed to increase.
It is usually added that the current uranium price does only have a negligible effect on the 
production cost of the kWhe. Arguments that, instead of the monetary prize, 
the energy return over energy invested value needs to be taken into account 
are usually dismissed with the statement that current energy extraction costs of uranium are small 
compared to other costs and the very low grade Rossing mine in Namibia is often given as 
a proof that one can still go a long way before the extraction energy cost will become 
important\cite{Rossingmine}. 

However, instead of exchanging endless arguments about the limits of this approach, we propose 
to use the Red Book uranium resource data base to test the above hypothesis, called in the following:

\begin{center}
\large{\bf The ``economic-geological hypothesis"}\\
\end{center}

This can be done easily as the Red Book quantifies the RAR and IR uranium resources according to 
almost equal and increasing cost intervals of about 40 dollar and 50 dollars. 
According to this hypothesis much larger uranium quantities are expected for the higher cost 
categories. 


\begin{center}
\large{\bf The ``economically extractable uranium resources are limited" hypothesis}\\
\end{center}

An alternative hypothesis assumes basically that uranium and its use in the energy sector 
has nothing special in comparison with any other energy resource. Consequently the ``law of diminishing return" 
applies also to uranium and the exploitation and use of uranium will follow these lines:

\begin{itemize}
\item The usage of uranium will start with the finding and exploitation of the 
big and high ore grade uranium deposits.
\item Once these big ``elephants" are hunted, one will turn to smaller and lower grade
uranium deposits. One tries to keep on going by developing and using better and better technology.
\item  Eventually the interesting 
deposits at the mine become too small and too diluted and the production will be terminated.
This moment is reached in a similar way for oil and for uranium, 
and according to the argument of M. K. Hubbert \cite{hubbertquote}
when he replied to David Nissen - Exxon: \\ 
{\it ``.. [T]here is a different and more fundamental cost that is independent of the monetary price.
That is the energy cost of exploration and production. So long as oil is used as a source of
energy, when the energy cost of recovering a barrel of oil becomes greater than the energy
content of the oil, production will cease no matter what the monetary price may be."}
\end{itemize}
 
While this hypothesis is theoretically very attractive, it can not be used 
easily to make a quantitative test. For example 
the energy extraction costs and the total energy costs 
are rarely given in the required detail. Furthermore, it is essentially impossible to quantify 
the potential ``next round technological improvements". For example 
breakthrough new reactor concepts, based on the fuel breeding concept and including perhaps the use of thorium, 
and much better, but so far unknown, uranium extraction techniques can always be 
imagined. Thus, as is the case with the peak oil hypothesis, most people will accept 
this idea only once an exact peak date for the nuclear energy  
can be determined. This, of course, can be done only some time after the final decline becomes obvious. 

\subsection{Evolution of Uranium Resources according to the Red Book}

\subsubsection{A consistency check of the NEA Press declaration} 
We now turn to two claims made in the Red Book 2007 edition (abbreviated in the following as RB07 where the number 
indicates the publication year) and transmitted by their own press declaration to 
the media (Quote)\cite{NEA2007press}: \\ 

\noindent
(1) {\it ``There is enough uranium known to exist to fuel the world's fleet of nuclear reactors at current consumption rates for at least a century, according to the latest edition of the world reference on uranium published today.
{\bf Uranium 2007: Resources, Production and Demand}, also known as the Red Book, estimates the identified amount of conventional uranium resources which can be mined for less than USD 130/kg* to be about 5.5 million tonnes, up from the 4.7 million tonnes reported in 2005."} \\

\noindent 
(2) {\it ``The currently identified resources are adequate to meet the expansion of nuclear power plants 
from 372 GWe in 2007 to between 509 GWe (+38\%) and 663 GWe (+80\%) by 2030."} \\

\noindent
Let us recalculate the numbers presented to the media. The yearly uranium needs to operate 
the existing nuclear power plants with the 2009 capacity of 370 GWe are about 65000 tons. 
As it is claimed above and quantified in the RB07, the 
conventional uranium resources of 5.5 million tons are the sum of the RAR ($<$ 130 dollar/kg), given as 3 338 300 tons,  
and the believed to exist IR 
($<$ 130 dollar/kg), given as 2 130 600 tons.
Following this logic, a simple division tell us that these 5.5 million tons of uranium resources, at constant usage, 
are sufficient at best for 85 years or "almost a century" and not ``for at least a century"! 

Furthermore, a more correct press declaration would perhaps say: \\

 {\it ``The well known RAR numbers
have remained roughly constant during the past years and these known resources 
are sufficient to operate the current world's reactor fleet for about 51 years only.  
However, since the amount of believed to exist IR resources has increased by about 700000 tons,
another 34 years can be added  
if all this IR uranium can indeed be extracted."} \\

Next we can ask how long the conventional uranium resources will 
last under the condition of a 38\% or 80\% growth scenario between 2007 and 2030. 

Given these growth assumptions, by the 
year 2030, the yearly natural uranium requirements would be between 
90000 tons/year and 115000 tons/year.

For simplicity one could assume that the above increase  
would be achieved with an average 23 year growth rate of 1.4\%/year 
and 2.5\%/year respectively.
Following this growth model during the 23 years from 2007 to 2030 between 
1.76 and 2.02 million tons would have been used already by the 
year 2030. By the year 2030 the world reactor fleet would 
need between 90000 tons/year and 115000 tons/year.
If one assumes the unlikely case that nuclear energy would remain constant 
after 2030, the claimed conventional uranium resources from 
2007 could thus fuel the 509 GWe power plants scenario 
up to the year 2071 and the 663 GWe scenario up to the year 2060.
Consequently, one finds that the operating lifetime of the reactors build during the 
years 2020 to 2030 will be limited by the amount of identified fuels and not by the 
expected 60 year lifetime. 

These simple examples show that the claims made in the NEA press declaration
are not justified by their own Red Book data. 

\subsubsection{A 20\% increase of the conventional uranium resources?}

As the reported increase of conventional uranium resources between 2005 and 2007 
is nevertheless relatively large, it might be interesting to learn where and in which price category the increase has happened.
Furthermore, one might be curious to see how the reduced dollar value and the increased mining costs 
are reflected in the pseudo-precise resource data and if a reduction from the yearly uranium extraction is included 
for the different countries. 

We first present, Tables 2-5, the world total resource estimates for the different categories 
and their evolution as given in the last 4 Red Book editions   
from 2001, 2003, 2005 and 2007 \cite{RB010305}.  
In order to simplify the discussion, the numbers are recalculated such that the uranium amounts for a given cost interval can be compared. Table 2 shows the evolution of the conventional resources 
since 2001. As one can see, the always highlighted huge increase is essentially only 
associated with some changes in the undiscovered but believed to exist IR resources.
Furthermore the presented RAR data do not indicate that the yearly uranium extraction 
of roughly 40000 tons has been taken into account.  
Table 3 and 4 show the corresponding evolutions for the RAR and IR categories 
split according to the estimated extraction cost range. 
 
\small{
\begin{table}[h]
\vspace{0.3cm}
\begin{center}
\begin{tabular}{|c|c|c|c|}
\hline
Red Book year      & RAR  [tons]                     & IR  [tons]             & conventional resources [tons] \\
                             & $<$ 130 dollars/kg          & $<$ 130 dollars/kg             & $<$ 130 dollars/kg   \\
\hline
2001                    & 2853000               &   1080000                           & 3933000  \\
2003                    & 3169238               &   1419450                           & 4588688  \\
2005                    & 3296689               &   1446164                           & 4742353  \\
2007                    & 3338300               &   2130600                           & 5468800  \\
\hline
\end{tabular}\vspace{0.1cm}
\caption{The evolution of the conventional uranium resources split into the 
reasonably assured resource (RAR) and the inferred resource (IR) categories from the latest four 
Red Book editions.  
Especially remarkable is fact that the RAR numbers have increased by only a small amount 
and remained essentially stable since 2003. Thus the claimed large increase in 
conventional uranium resources since 2001 and especially during the  past 4 years is 
only based on the increased IR number.}
\end{center}
\end{table}
}

\small{
\begin{table}[h]
\vspace{0.3cm}
\begin{center}
\begin{tabular}{|c|c|c|c|}
\hline
Red Book year      & RAR  [tons]                                    & RAR  [tons]                           & RAR [tons] \\
                             & $<$ 40 dollars/kg          & 40-80 dollars/kg             & 80-130 dollars/kg   \\
\hline
2001                    & 1534100               &   556650                             & 589770  \\
2003                    & 1730495               &   575197                             & 661941  \\
2005                    & 1947383               &   695960                             & 653346  \\
2007                    & 1766400               &   831600                             & 740300  \\
\hline
\end{tabular}\vspace{0.1cm}
\caption{The evolution of the reasonably assured resource (RAR) category from the latest four 
Red Book editions.  Especially remarkable is that the highest uranium numbers are found in the lowest cost category 
and this category has, after regular large increases, suddenly decreased since 2005 by about 180000 tons.}
\end{center}
\end{table}
}

The RAR numbers, even though claimed to be known with unbelievable precision, 
appear to fluctuate by a large amount. 
The drop of 180000 tons in the cheapest and best understood $<$ 40 dollars/kg category 
between 2005 and 2007 is certainly remarkable and more details about this reduction will be given in section 3.

Otherwise and on first view, ups and downs of about $\pm$ 10\% appear perhaps to be reasonable.
For example one might expect that inflation moves some resources from the cheaper to 
the more expensive ones. Such an explanation requires also that 
a certain amount from the highest cost category becomes 
out of scale. 

Next we turn to the more and more speculative uranium resources.
In Table 4 the not yet found but believed to exist 
IR uranium data are presented.  Especially suspicious is the large increase of 400000 tons 
in the $<$ 40 dollars/kg IR category. This increase can be compared with the corresponding RAR numbers from Table 3,  
which decreased during the same period by 180000 tons. 

The situation becomes even more bizarre when one compares the 
evolution of the IR category during the past 6 years from 2001 to 2007. 
For example the 40 dollars/kg IR category increased by a factor of 2.2 and the 40-80 dollars/kg category by a factor of 3.5. 
In comparison the amount in the 80-130 dollars/kg category changed only 
by a factor of 1.3. 
Finally one can compare the evolution of the conventional 
resources in the RAR category and the more speculative IR category. 
As mentioned already, large exploration efforts during the past years 
have left the total RAR numbers essentially unchanged but have 
increased the believed to exist IR figure by a large amount.
This means that the claimed increase from 2005 to 2007 in the conventional 
uranium resources is not based on real discoveries, but on an 
unexplained hope factor associated with the IR deposits, eventually to be discovered.

More details about these changes will be discussed in the individual country analysis below. 
    
\small{
\begin{table}[h]
\vspace{0.3cm}
\begin{center}
\begin{tabular}{|c|c|c|c|}
\hline

Red Book year      & IR  [tons]                                    & IR     [tons]                        & IR  [tons]\\
                                & less than 40 dollars/kg          & 40-80 dollars/kg             & 80-130 dollars/kg   \\
\hline
2001                    &  552000               &    186950                             & 225150  \\
2003                    &  792782               &    275170                             & 320868  \\
2005                    &  798997               &    362041                             & 285126  \\
2007                    & 1203600               &   655480                             & 272200 \\
\hline

\end{tabular}\vspace{0.1cm}
\caption{The evolution of the not yet discovered but believed to exist IR uranium resources 
as given in the last four editions of the Red Book. Remarkable is the claim that 
the cheaper cost categories increased by a large amount but the highest cost category has even decreased.}
\end{center}
\end{table}
}

Table 5 shows the evolution for the undiscovered prognosticated and speculative UPR and USR
resource categories. In contrast to the increase from 2003 to 2007 in the conventional IR resources, 
only relatively minor changes are claimed for the even more uncertain UPR and USR resources. 

\small{
\begin{table}[h]
\vspace{0.3cm}
\begin{center}
\begin{tabular}{|c|c|c|c|}
\hline

Red Book year      & UPR    [tons]                                     & UPR  [tons]                           & USR   [tons]                                      \\
                                & less than 80 dollars/kg          & 80-130 dollars/kg             & less than 130 dollars/kg          \\
\hline
2001                    & 1480000               &    852000                             & 4438000  \\
2003                    & 1474600               &    779900                             & 4437300  \\
2005                    & 1700100               &    818700                             & 4557300  \\
2007                    & 1946200               &    822800                             & 4797800 \\
\hline
\end{tabular}\vspace{0.1cm}
\caption{The evolution of the undiscovered prognosticated UPR and speculative USR uranium resources 
according to the past four Red Book editions. In comparison to the large relative changes in the IR data 
the numbers presented show an astonishing stability.}
\end{center}
\end{table}
}

One finds from Tables 3-5 that the uranium resources in the RAR, IR and 
UPR categories decrease for the higher cost intervals. 
Furthermore,  one observes  
that the estimated world RAR as well as the IR and UPR numbers have changed in some very particular and unnatural ways.

{\bf Thus, the overall uranium resource data and their evolution are in contradiction with the ``economic-geological 
hypothesis"}, presented in section 2.2.

Furthermore, and if inflation effects are ignored, 
one would expect that the changes of the uranium quantities  
in the different cost RAR, IR and UPR categories for the RAR, the IR and the UPR should follow similar trends.
As the uranium resource data do not confirm such expectations, one is left with the conclusion
that the pretended high quality uranium data do not exist.

\section{Evolution of uranium resources in selected countries}

In order to understand how and where uranium resources have changed  
during the past few years, one needs to study the information provided 
by the correspondents from a few different countries with large resources. 
For this purpose the Red Book editions from the years 2003 (RB03), 2005 (RB05)and 2007 (RB07) will be used.
We restrict the discussion to the 10 countries, which claim to have more 
than 100000 tons of extractable RAR uranium resources for $<$ 130 dollars/kg within their territory.
Combined, these 10 countries, cover a surface of about 52 million km$^{2}$ or 
more than 1/3 of the total land surface of our planet. 
After at least 50 years of non negligible worldwide geological research efforts, 
these countries claim to have 80\% of the remaining known world uranium resources and up to 95\% of the uranium 
in the economically most interesting $<$ 40 dollars/kg RAR category and
roughly 90\% of the total uranium extraction in 2007 came from these countries.

Table 6 and 7 show the claimed amount of RAR uranium resources for these countries in the $<$ 40 dollars/kg  
and 40-130 dollars/kg categories.
   
\small{
\begin{table}[h]
\vspace{0.1cm}
\begin{center}
\begin{tabular}{|c|c|c|c|}
\hline
country      &  RAR (RB03)      & RAR (RB05)                             & RAR (RB07) \\
                  &  40 dollars/kg  [tons]    &     40 dollars/kg  [tons]                          & 40 dollars/kg  [tons]  \\
\hline

Australia                    & 689000               &   701000                             & 709000  \\
Brazil                    &      26235               &   139900                                  & 139600  \\
Canada                   & 297264               &   287200                               & 270100     \\
Kazakhstan                   & 280620               &   278840                             & 235500  \\
Namibia                   &       57262               &   62186                             & 56000  \\
Niger                   & 89800               &   172866                             & 21300  \\
Russia                   & 52610               &   57530                             & 47500  \\
South Africa                   & 119184               &   88548                             & 114900  \\
Ukraine                   & 15380               &   28005                             & 27400  \\
USA*                   & 102000               &   102000                             & 99000  \\
sum                   &   1627000                            &    1714000                                         & 1621000              \\
\hline
\end{tabular}\vspace{0.1cm}
\caption{Evolution of the low cost RAR uranium category for 10 countries which claim to have 
a total of more than 100000 tons of RAR resources on their territory. An especially remarkable 
change during the years 2005 to 2007 can be seen for Niger. 
$^{*}$The USA report do not report a number for the $<$ 40 dollars/kg RAR
category, the amount in the $<$ 80 dollars/kg is used here.}
\end{center}
\end{table}
}

\small{
\begin{table}[h]
\begin{center}
\begin{tabular}{|c|c|c|c|}
\hline

country      &  RAR (RB03)      & RAR (RB05)                                                            & RAR (RB07) \\
                  &  40-130 dollars/kg  [tons]    &     40-130 dollars/kg  [tons]                          & 40-130 dollars/kg  [tons]  \\
\hline

Australia                    & 46000               &   46000                             & 16000  \\
Brazil                    &      59955               &     17800                                  & 17800  \\
Canada                   &   36570               &      58000                               & 59100     \\
Kazakhstan                   & 249840               &   235057                             & 142600  \\
Namibia                   &       113270               &   120370                             & 120000  \\
Niger                   & 12427               &   7600                             & 222180  \\
Russia                   & 90410               &   74220                             & 124900  \\
South Africa                   & 196146               &   167045                             & 169500  \\
Ukraine                   & 49280               &   38701                             & 107600  \\
USA*                   & 345000               &   342000                             & 339000  \\
sum                   &   1198900                            &    1106800                                         &  1318700             \\
\hline                 
\end{tabular}\vspace{0.1cm}
\caption{Evolution of the higher cost RAR uranium category for 10 countries which claim to have 
a total of more than 100000 tons of RAR resources on their territory. Especially remarkable 
are the changes from 2005 to 2007 for Australia, Kazakhstan, Niger, Russia and the Ukraine. These changes 
in comparison with the ones in the  
low cost category presented in Table 6 are also interesting.$^{*}$As the USA does not report the $<$ 40 dollars/kg RAR
category, the amount in the 80-130 dollars/kg category is used.}
\end{center}
\end{table}
}

\small{
\begin{table}[h]
\begin{center}
\begin{tabular}{|c|c|c|c|}
\hline

country      &  ratio (RB07/RB05)                         & ratio (RB07/RB05)               & ratio RB07/RB05 \\
                  &  IR $<$ 40  dollar/kg  [\%]    &     IR 40-130 dollar/kg  [\%]       & UPR $<$ 130 dollar/kg  [\%]  \\
\hline
Australia                    & 1.42               &   0.48                             & NA  \\
Brazil                    &      1               &     1                                  & 1  \\
Canada                   &   0.97              &      0.83                              & 1     \\
Kazakhstan                   & 2.18               &   0.91                             & 0.97  \\
Namibia                   &       0.99                &   0.99                             & NA  \\
Niger                   &  NA               &   0.4                             & 1   \\
Russia                   & 1.67               &   17.7                             & 2.65  \\
South Africa                   & 2.19               &   1.01                             & 1  \\
Ukraine                   & 1.03               &   3.5                            & 1.47  \\
USA                   &  NA               &   NA                             & 1  \\
world total         &   1.5                            &    1.43                                        &  1.1            \\
\hline                 
\end{tabular}\vspace{0.1cm}
\caption{The IR resource ratios as obtained from the Red Book 2007 and 2005 editions and for the 
10 countries which claim to have a total of more than 100000 tons of RAR resources on their territory. 
Not all countries have submitted or updated these numbers for the 2007 edition. 
Especially remarkable changes are observed for Russia  where the category IR (40-130 dollar/kg) is now estimated to be 337000 tons. Some changes for Australia, Kazakhstan, Niger and the Ukraine are also interesting.}
\end{center}
\end{table}
}

Some spectacular ups and downs can be observed for the three Red Book editions. For example,
between 2005 and 2007, the RAR reserves    
in the $<$ 40 dollars/kg category decreased by 15\% (minus 40000 tons) for Kazakhstan and by 88\%
(minus 150000 tons) for Niger. 
Drastic changes during these two years are also reported in the 40-130 dollars/kg RAR category
for Australia, Kazakhstan, Niger, Russia and the Ukraine.
Despite the fact that the RAR numbers, especially in the less than 40 dollars/kg category, 
are assumed to present the most accurate estimate, essentially no explanations for the often dramatic changes are given. 

The changes for the yet {\bf unobserved} but believed to exist IR resources are sometimes even more interesting.
As presented in section 2, and in contrast to the essentially unchanged claimed total RAR resources, 
the data reported for the IR category and the $<$ 130 dollars/kg price tag 
have increased by almost 700000 tons between the year 2005 and 2007. 
The sometimes spectacular and unlikely relative changes for some countries can be seen from 
Table 8 where we present ratios of the resource numbers found presented in RB07 and RB05 and 
for two IR cost categories and for the UPR category.
An especially remarkable 
increase is observed for Russia. It is claimed that their IR 40-130 dollars/kg category increased by a factor of 17.7 
from 19000 tons to 337000 tons. 
The reported changes of the IR data for Australia, Kazakhstan, Niger and the Ukraine are also interesting. \\

As we have seen already in section 2, the celebrated increase of the conventional uranium resources 
does not come from new discoveries of interesting uranium deposits, but from a new evaluation of the 
supposed to exist IR resources. This statement can now be made more accurately! 
The data show that this claimed increase of the IR resource 
comes essentially only from Russia (from 40652 tons to 373300 tons),
Australia (from 396000 tons to 518000 tons) , Kazakhstan (from 302202 tons to 439200 tons) 
and the Ukraine (from 23130 tons to 64500 tons). 

A closer look at Russia shows that this increase is very suspicious. While 
the IR number in the $<$ 40 dollars/kg category changed by only 15000 tons from 
21572 tons to 36100 tons, an incredible increase from 19080 tons to 337200 tons is presented 
for the 40-130 dollars/kg category. 

Kazakhstan is another particular example for drastic changes of the IR data. 
From the RB05 and RB07 one finds that the $<$ 40 dollars/kg category IR number for Kazakhstan increased 
from 129252 tons in the 2005 estimate to 281800 tons in 2007. 
The 40-130 dollars/kg number decreased however from 172950 to 157400 tons.
In comparison, the very speculative UPR and UPS data for Kazakhstan remain essentially unchanged.

As discussed in Chapter I and II of this report\cite{chapter12}, the evolution of  uranium mining in Kazakhstan  
is of particular importance to avoid a world uranium supply shortage during the coming 5-10 years.
It is claimed, provided that enough investments in the mining are done, that this country can triple uranium extraction within the next 10-15 years 
from 6637 tons in 2007 to 21000 tons in 2015. A very large number for future uranium discoveries in the low cost category will certainly help to raise foreign interest for investments 
in the uranium mining infrastructure.   

%

Australia and South Africa also claim large increases, but their resources increased only in the 40 dollars/kg IR category. 
In contrast, the IR data for Canada, Brazil and the USA remained unchanged at their 2005 values.

The above examples demonstrate that a large amount of the uranium resources and 
their evolution are not obtained from geological methods.

\subsection{Are some uranium resource data not based on geological methods?}

If one accepts that uranium resource data for many countries are 
not based on geological methods,
it follows that other methods have helped to fill the tables of the Red Book. 

Consequently, and in absence of explanations, one is somehow invited to formulate some ideas 
about why some particular countries, probably with the help from large 
mining companies, might be interested in presenting too high or too low resource numbers.  

For example one can imagine that "sudden" increases in resource numbers, 
as observed for Australia, Kazakhstan, Russia and South Africa, will help to attract foreign investment.  

On the contrary, a sudden and drastic reduction in the most interesting $<$ 40 dollars/kg RAR category, 
as observed for Niger, could be motivated by ideas (1) that this keeps potential uranium mining competitors 
out of the country or (2) that some companies are not interested to inform the government and the people 
about their particular richness in mineral resources. 

\subsection{Relations between the different cost categories}

We now compare the individual country resource data with the 
``economic-geological hypothesis" presented in section 2.2. 
Starting with the lowest and highest RAR and IR cost categories of $<$ 40 dollars/kg and 80-130 dollars/kg, 
one finds that some country estimates show surprising large differences in these categories with respect to the 
world average. For example 53\% of the word RAR resources are expected in the 
$<$ 40 dollars/kg categories but only 22\% in the 80-130 dollars/kg category and in 
disagreement with the ``economic-geological hypothesis".

The disagreement with this hypothesis becomes even larger for Australia, Canada and Kazakhstan. 
For Australia one finds that 98\% of the RAR is claimed to be 
in the low cost category. Too high numbers for this category are also reported from 
Canada (82\%) and Kazakhstan (62\%). 

In contrast, the numbers in this cost category from Russia (28\%) and 
for Niger (9\%)\footnote{For Niger the uranium amount in this class is now given as 21300 tons, which is about 
150000 tons smaller than in the 
the 2005 edition when 96\% of the country's RAR resources were assigned to the $<$ 40 dollars/kg category.}
are very low in comparison. 

The data reported for the IR category show similar discrepancies between the world average ratios and the ones from individual countries. One finds that 56\% and 13\% of the IR resources are predicted respectively in the 
$<$ 40 dollars/kg and in the 80-130 dollars/kg categories.
In comparison, the correspondents from Australia, Canada and Kazakhstan think that 94\%, 88\% 
and 64\% respectively will be found in the 
40 dollars/kg category. The three countries thus predict that their not yet discovered IR resource fractions 
match almost perfectly the corresponding RAR fractions. 

In contrast, the correspondents from Russia 
assume that only 9.7\% of their IR will eventually be found in this low cost category. For Niger this IR fraction is 
given as 42\% and thus very close to the world average.

{\bf The Red Book uranium resource data show that the``economic-geological hypothesis" is not backed up by the data.
This conclusion is strengthened beyond doubt if one believes 
that Australia and Canada provide the most reliable resource data.}

The relation between the RAR numbers and the IR numbers  
is also interesting. For the $<$ 40 dollars/kg category, Australia assumes to know about 709000 tons RAR 
and expects to find another 487000 tons in the IR category,  or 69\% of the RAR number. 
In contrast for Canada the RAR number, given as 270000 tons, the IR number is 82000 tons, thus only 30\% of the 
RAR number.

\subsection{Uranium mining and its effect on resource data?}  

Finally, we would like to see how uranium extraction, claimed to be known accurately to the 
ton, e.g. far better than a 0.1\% accuracy, influences the remaining amount of uranium  
in the different RAR resource categories and in some selected countries. 

For this investigation we remind the reader that worldwide about 40000 tons of uranium are mined 
on average per year. For many years, and despite non-negligible efforts by many countries, 
only three countries extract about 60\% of this uranium and  
individually more than 5000 tons per year. Another 25\% of this uranium comes from three countries which 
contribute each about 3000 tons/year and further 12\% from three additional countries which 
extract together roughly 5000 tons/year. 

Furthermore, the uranium extraction is concentrated within the hands of a few transnational mining companies.
For example the four biggest, Rio Tinto, Cameco, Areva and KazAtomProm
provided about 26000 tons/year (about 59\% in 2008) to the world uranium market. 
Despite the claim that plenty of cheaply extractable uranium 
can be found almost everywhere on the planet and that 
the extraction cost does not play a role, 66\% of the 41000 tons extracted in 2007 came from only 
10 uranium mines.

The biggest mine today, McArthur River in Canada owned dominantly by Cameco,  
extracted 7200 tons of uranium in 2007, or about 18\% of the worldwide production.

This number might be compared with today's stressed world oil situation, where the largest oil field ever, Ghawar in Saudi Arabia contributes about 6\% of the total world oil production.
In fact it might be better to compare the fraction of uranium production 
from this mine alone with the fraction of oil produced from Saudi Arabia and Kuwait combined.
 
The mine started only about 10 years ago and reached 7200 tons/year during the years 2002-2007. 
Since the start in the year 2000, about 58000 tons have been extracted.  
According to Cameco this mine exploits the world largest high-grade uranium deposit 
with proven and probable reserves of 332.6 million pounds of $U_{3}O_{8}$. Thus an equivalent of more than 
almost 130000 tons of natural uranium, with about 65000 tons assigned (31.12.2008) as proven 
reserves\cite{McArthurmine}. This mine seems to be past peak as 
in 2008 only 6383 tons were produced and the output from the first half of 2009 
reported by Cameco on August 12, \cite{Cameco2009}, appears to be again some 12\% lower than the one obtained 
during the same period in 2008. If one assumes that about 50\% of the extractable uranium 
had been extracted up to 2005/06, the presumed peak year, one could estimate that 
instead of the 65000 tons only about 45000-50000 tons remain to be mined. 
The next few years will tell if the decline observed since 2007 will continue.   

The next two mines, Ranger in Australia and Rossing in Namibia, produced
together 8000 tons of uranium in 2008,  or only about 25\% more than the McArthur River mine. 
Combined the three largest mines produced 33\% of the total and thus 
slightly more uranium than the next 7 big uranium mines. 
This fraction corresponds roughly to the entire OPEC share of the world oil production.

Thus, uranium extraction is much more centralized and monopolized than any other 
energy resource. In fact, if the world oil situation (with a few giant oil companies and 
a country cartel) frightens policy makers and most oil consumers,
the uranium situation is by all standards much more dangerous.  \\

We now try to see if the amount of uranium extracted during the past years
has some effect on the RAR numbers.
For this study we use the uranium quantities extracted during the past few years 
as provided from the different editions of the Red Book and repeated 
in a simple table in a WNA information paper\cite{WNAminingpaper}.

Starting with the largest producer country Canada, one finds that the three large existing mines extracted  
essentially 100\% of the 9477 tons and 9000 tons in 2007 and 2008. During the years 2003-2004 and 
2005-2006 the total extracted uranium is given as 22055 tons and 21491 tons respectively.
Table 6 shows that the $<$ 40 dollar/kg RAR category decreased 
during these two year periods by 10064 tons and 17100 tons respectively. 
As it seems most likely that the existing big uranium mines operate and deplete only the 
$<$40 dollar/kg category the numbers show that only 50\% (2005) and 80\% (2007) of the decrease 
can be accounted for directly. Two explanations are possible, (1)
about 12000 tons (2003 + 2004) and 4000 tons (2005 + 2006) of new deposit in the $<$ 40 dollar/kg RAR category have been discovered during the considered two year periods or (2) extraction figures are not taken into account. 

We now turn to Australia, the second world contributor of uranium. During the four years (2003-2006) 
a total uranium extraction of 33663 tons is reported and the 
$<$ 40 dollar/kg category increased by 20000 tons (Table 6) 
As Australia does not claim to have significant amounts of uranium in the 40-80 dollar/kg 
and 80-130 dollar/kg RAR categories, one finds again that essentially all of the extracted uranium came 
from the 40 dollar/kg RAR category. 

Consequently the new findings in this category, and between 2003 and 2006, 
must have been about 54000 tons. 
However, such large new uranium discoveries over a four year period are somehow puzzling
as the other two RAR cost categories 40-80 dollar/kg and 80-130 dollar/kg remained 
unchanged between 2003 and 2005 and even decreased 
by 8000 tons and 22000 tons between 2005 and 2007.   
Thus, the extraction numbers from Australia are clearly inconsistent with the reported RAR numbers. 

As a last example we analyze the situation in Niger, a former french colony which became independent 
in 1960.  It is one of the poorest countries in the world with an electricity production of roughly 0.234 billion kWh (2005),
corresponding to an almost negligible 18 kWh per year and per person. 
Yet, the 3032 tons of uranium extracted in 2008 allowed to fuel almost 20 GWe nuclear power plants in France and Western Europe, which produced roughly 140 billion kWh during that year.
Between 2003 and 2006 about 13000 tons of uranium have been extracted
from the mines operated dominantly by AREVA, a french transnational 
nuclear company. 

In 2003, the RAR resources were reported as 89800 tons in the 
$<$ 40 dollars/kg and 12447 tons in the 40-130 dollars/kg category. 
These numbers changed by incredible amounts to 172866 tons and 7600 tons respectively in 2005. 
Another drastic change is reported in the 2007 Red Book  
and the corresponding RAR numbers are now given as 21300 tons and 222180 tons respectively.  

Obviously, neither the 13000 tons of uranium extracted during these 4 years are accounted for, nor 
does anybody of the Red Book authors seem to be surprised about 
the incredibly large jumps for the $<$ 40 dollars/kg and 40-130 dollars/kg RAR categories.
 
These numbers must thus contain some fantasy factor, which can perhaps be explained with   
the misinformation hypothesis.  This is further supported by AREVA's problems with the real owners,
often called ``Tuareg rebells", who ask for a larger share in the profits. 

In summary, the claimed high precision uranium resource data and the known extraction data from the past few years  
do not match. These and the other inconsistencies described in sections 2 and 3 
raise suspicions about the accuracy of the RAR uranium data.

\section{Consequences for the long term nuclear energy future.}

The analysis, presented in section 2 and 3, demonstrates that the uranium 
resource data, prepared, updated and published every two years by the IAEA and the NEA in the Red Book, 
do not measure up to the claimed high precision standards.
On the contrary, it even seems that some individual country resource data 
are not based on a scientific geological resource estimate.

Consequently some large error margins even for the ``reasonable assured resources" (RAR)
category should be used. 
As an example one could assume that the RAR resource numbers from Australia 
and Canada are known best and use their own estimate that 
only the amount in the $<$ 40 dollars/kg category is relevant for the mining 
with today's mining technology. If this idea is applied to the entire world 
one would guess that only 
$<$ 40 dollar/kg RAR category are exploitable.
As a result, the known uranium resources could be guessed as $\leq$ 2 million tons, 
corresponding to a static resource lifetime of just 30 years.

Such an evaluation would certainly discourage the idea to construct new 
standard light water reactors with a presumed lifetime of 60 years.

This simple-minded example demonstrates that realistic uranium resource information is 
urgently needed. Such an analysis, clearly beyond the scope of this paper,  
would be based on a critical mine by mine and country 
by country analysis. 

Despite these shortcomings, the Red Book uranium resource data 
are the only existing and usable data base. 
These data, including large uncertainties, demonstrate that 
the ``economic-geological hypothesis" is contradicted by the data. This widely used hypothesis 
states that more and more uranium can be extracted  
if only the price is allowed to go up. This claim is in total disagreement with the overall resource data and with the ones 
from many individual countries.  
   
{\bf Thus, one is left with the choice of either rejecting the Red Book data completely 
and sticking with an unproven hypothesis, or giving up that unproven hypothesis.} \\

In summary we point out that countries 
interested in the construction of a new nuclear power plant within the next 10-20 years
should find a way that their uranium fuel can be guaranteed at least for 40 years before  
they invest perhaps up to 4 billion Euro per GWe installed power.    

The warning applies also to basically all Western European countries, Japan and South-Korea which depend 
to almost 100\% on stable uranium deliveries from far away. 
These countries should take one particular paragraph from the Red Book 2007 NEA press declaration very seriously:

{\it ``At the end of 2006, world uranium production (39 603 tonnes) provided about 60\% 
of world reactor requirements (66 500 tonnes) for the 435 commercial nuclear reactors in operation. 
The gap between production and requirements was made up by secondary sources draw down from government 
and commercial inventories (such as the dismantling of over 12 000 nuclear warheads and the re-enrichment of uranium tails). 
Most secondary resources are now in decline and the gap will increasingly need to be 
closed by new production. Given the long lead time typically required to bring new resources into 
production, uranium supply shortfalls could develop if production facilities are not implemented in a timely manner."}

Many other reports have studied the world uranium supply situation in detail.
Even though most of these reports assume, contrary to our study, 
that the Red Book uranium resource data are roughly 
correct, very similar conclusions about the short and long term critical uranium supply situation 
are reached. 
The list below provides references to some recent studies which 
reach the conclusion that the known uranium deposits and techniques of uranium extraction 
are not sufficient to fuel a nuclear energy renaissance based on conventional light water reactors. 

The following three studies are from groups that favor nuclear energy. They find that 
even a small 1\% annual nuclear power growth scenario 
will be faced with serious and unsolved uranium supply problem during the first half of the 21st century.

\begin{itemize} 
\item The report from the year 2002, ``A Technological Roadmap for Generation IV Nuclear Energy 
Systems" \cite{GenIV} points out that the known conventional uranium resources 
will only last between 30-50 years. Thus, a new conventional nuclear power plant which might be operational 
in 2020 might only obtain uranium fuel up to sometime between 2040 and 2050. 
\item  The IAEA 2001 report ``Analysis of uranium supply to 2050" \cite{uran2050}. 
The authors of this report quantify the uranium deficit with respect to the RAR numbers 
and for different scenarios about the future use of nuclear fission energy.  The estimated deficit 
is given in units of million of tons of uranium. Many details about the potential contributions of uranium 
from a large number of unconventional resources are presented in this report (section 5) and 
especially the remarks about sea water uranium are remarkable (quote): 

{\it ``Research on extracting uranium from sea water will
undoubtedly continue, but at the current costs sea water
as a potential commercial source of uranium is little
more than a curiosity."}
\item A 2007 MIT study group concluded that \cite{MIT2007}
``Lack of fuel may limit U.S. nuclear power expansion". 
\end{itemize}

Another group of analysts, with critical views about nuclear fission energy,  
have also studied the Red Book uranium resource data. 
All these studies, even if they assume that the Red Book 
uranium resource numbers are more or less accurate, conclude that a substantial increase of 
nuclear fission energy, using conventional light water reactors, is essentially impossible.

\begin{itemize} 
\item The ``Energy Watch Group" report from December 2006\cite{Zitteluran}
with Dr. Werner Zittel and J\"org Schindler from the Ludwig B\"olkow Systemtechnik GmbH
as the principle authors conclude that (quote): \\
 
{\it ``If only 42000 tons/year of the proved reserves below 40 dollar/kg can be converted into production
volumes, then supply problems are likely even before 2020. If all estimated known resources
up to 130 dollar/kg extraction cost can be converted into production volumes, a shortage can at
best be delayed until about 2050."} 
\item WISE Uranium Project\cite{wisetalk}
``Uranium supply and demand". Some interesting graphics, relating  
the various resource categories from the 2005 Red Book with some modest nuclear growth scenarios,
demonstrate the year when the uranium supply cliff will be reached.   
\item In the article ``The Red Face book" published at Sanders research (September 2008), 
John Busby has analyzed the 2007 Red Book in many details \cite{JohnBusby}. 
Many of the internal inconsistencies of the Red Book 2007
have been pointed out in this article and most likely for the first time. His presented conclusions, 
about the near and long term uranium supply troubles,
are essentially identical to the ones, obtained independently and with a somewhat different approach 
in the three chapters of this report \cite{chapter12}.
\item Another important report from Nov. 2007, ``THE LEAN GUIDE TO NUCLEAR ENERGY", by 
David Fleming \cite{Fleming}
has focussed on many issues of nuclear energy and its inconsistencies. D. Fleming 
concludes his studies with the statement (quote): \\
{\it ``Shortages of uranium Ð and the lack of realistic alternatives Ð
leading to interruptions in supply, can be expected to start in the
middle years of the decade 2010-2019, and to deepen thereafter."}
\item Finally we would like to reference the report,   
``Nuclear power the energy balance", by Jan Willem Storm van Leeuwen and Philip Smith 
and its latest update by Jan Willem Storm van Leeuwen \cite{Storm}.

This report tries to give, among other things, an energy balance of the entire nuclear power chain, starting  
from the mining to the waste disposal. It presents 
the hypothesis that ````Economically extractable uranium resources are limited". 
\end{itemize}

\section{Summary}

Despite the shortcomings of the Red Book and the associated large uncertainties, 
some valuable information can still be extracted.
Perhaps the most important results of our analysis are:
\begin{itemize}
\item The ``economic-gelogical hypothesis" that more uranium resources can be extracted 
if one is only willing to pay a higher price is in contradiction with  
the Red Book resource data. 
\item Realistic uranium resource data can not be obtained directly from the Red Book.
However, a detailed comparison of the data from current and past Red Books and the often far too drastic 
resource changes, following some observations from this analysis, 
can perhaps be used to obtain eventually a better resource estimate. 
\item The extractable uranium resources in many countries are most likely much smaller than generally 
believed. In absence of a Red Book document that measures up to its claims,  perhaps only the 
RAR uranium in the $<$ 40 dollars/kg category might be considered as realistic. 
\end{itemize}

The analysis, presented in this and the previous two chapters\cite{chapter12}, demonstrates 
that the current uranium extraction and the believed-to-exist uranium resources are incompatible 
even with a modest growth scenario of conventional nuclear fission 
power\footnote{Those who disagree with this conclusion 
are encouraged to present their own uranium resource analysis.}.

A debate about the future of nuclear energy must therefore be based on 
the two questions: (1) When -- if ever -- will reliable and safe commercial 
breeder reactors based on uranium or thorium become available? And (2) Will nuclear fusion power always be 50 years away.
The current situation and the prospects about these future hypothetical options will be presented 
in the last chapter IV of this report. 

It seem that our analysis could be best summarized with an addition to 
the recent warning from Faith Birol, the chief economist of the international energy agency\cite{Birol}:
{\it ``We should leave oil before it leaves us"}, stating that in fact 

{\it \bf ``We should also terminate the use of nuclear fission energy in the standard light water reactors 
before uranium leaves us as well"}.
      

\end{document}